\documentclass[prl,twocolumn,showpacs,]{revtex4}

\usepackage{graphicx,graphics,color,epsfig}
\usepackage{bm}
\usepackage{amsmath}
\usepackage{amssymb}
\usepackage{euscript}
\usepackage{longtable}

\newcommand{\beq}{\begin{equation}}
\newcommand{\eeq}{\end{equation}}
\newcommand{\beqn}{\begin{eqnarray}}
\newcommand{\eeqn}{\end{eqnarray}}

\begin{document}

\title{ Extended Recursion in Operator Space (EROS), a new impurity solver for the single impurity Anderson model}
\author{Jean-Pierre Julien}
\affiliation{Theoretical Division, Los Alamos National Laboratory, MS B262,  Los Alamos, NM 87545, USA}
\affiliation{Institut Neel CNRS and Universit\'{e} J. Fourier 25 Avenue des Martyrs, BP 166, F-38042
Grenoble Cedex 9, France}

\author{R. C. Albers}
\affiliation{Theoretical Division, Los Alamos National Laboratory, MS B262,  Los Alamos, NM 87545, USA }

\date{\today}

\begin{abstract}
We have developed a new efficient and accurate impurity solver for the single impurity Anderson
model (SIAM), which is based on a non-perturbative recursion
technique in a space of operators and involves
expanding the self-energy as a continued fraction. The method
has no special occupation number or temperature restrictions; the
only approximation is the number of levels of the continued
fraction retained in the expansion. We also show how this approach can
be used as a new approach to
Dynamical Mean Field Theory (DMTF) and illustrate this
with the Hubbard model. The three lowest orders of recursion
give the Hartree-Fock, Hubbard I, and Hubbard III approximations.
A higher level of recursion is able to reproduce the expected 3-peak structure
in the spectral function and Fermi liquid behavior.
\end{abstract}

 \maketitle

In the last decade, Dynamical Mean Field Theory (DMFT)
\cite{Georges96, Held07} has become a widely used method to
study strongly correlated electrons systems. It can be formulated
by an action formalism with a self-consistent mapping onto a single impurity
Anderson model (SIAM). Because the
critical step in this method is the quality of the impurity solvers,
there has recently been an increased interest
in new and more accurate SIAM solvers.
Currently available solvers include: iterated perturbation theory(IPT) \cite{IPT}, which
works well for small U and single band,  the non-crossing
approximation(NCA) \cite{NCA}, which is able to give the coherent peak  but
fails to reproduce Fermi liquid behavior, and the equation of motion
(EOM) method \cite{EOM}, which requires a decoupling scheme
and misses the Kondo peak at finite $U$ for
particle-hole symmetric systems. IPT also presents pathologies away from
half-filling. A numerical renormalization group (NRG)\cite{NRG} has
also been been employed but has the limitation that it captures correctly the low
energy physics but with a logarithmic divergence instead of the
expected Lorentzian shape of a Fermi liquid. It is also inaccurate
for high energy physics and is limited at T=0 though finite T is
tried to be implemented \cite{NRG-T}. Quantum Monte Carlo (QMC)
\cite{QMC}, as a non-perturbative approach, is in principle
rigorous  and provides good results at high temperature for static
properties but introduces uncertainty for dynamical properties
(like spectral density) because of the poorly defined inverse
problem, and cannot reach zero temperature.

The purpose of an Extended Recursion method in Operator Space
(EROS) is to provide an efficient (rapid) and accurate method to
solve the quantum impurity problem. Efficiency is especially
important for LDA+DMFT calculations, since most of
the computational time is spent on the quantum impurity problem.
An accurate solution must include a reliable calculation for both low and
high temperatures, low and high-energy physics, as well as for any
filling factor for the correlated
bands. We also intend to extend this method to quantum
transport problems for low-dimensional correlated systems (e.g., quantum
dots and molecular electronics). As shown below, EROS is able to retrieve well-known
approximations such as Hartree-Fock, Hubbard I, and Hubbard III, at the lowest level
of recursion and the
Fermi liquid regime and metal-insulator transition at higher levels.

The solution involves calculating a retarded Green function
 $G(\omega)\equiv <<A;B>>_{\omega}$, which is the Fourier
transform of $G(t)=-i\theta(t)<\{A(t),B^{\dag}(0)\}>$ for the operators
A and B.  This can be done through a recursion process (RP)
(see \cite{MoriZwanzig,FuldeBook}),
which can be directly seen by examining the coupled equations of motion
for the retarded Green function for a Hamiltonian $H$:
\beqn
<<A;B>>_{\omega}=\frac{<\{A,B^{\dag}\}>}{\omega} \nonumber\\
+\frac{<\{[A,H],B^{\dag}\}>}{\omega^2}+
\frac{<\{[[A,H],H],B^{\dag}\}>}{\omega^3}...
\label{momentexp}\eeqn This can be identified with a
moment expansion through
the definition \beq
 \mu_n= <\{[...[A,H],H]..,H],B^{\dag}\}>, \eeq
 where $H$ appearing $n$ times introduces a \textit{super}-operator
$\mathcal{H}$, the 'Liouvillian', acting in
operator space: \beq A\mathcal{H}=[A,H] \label{liouville}\eeq
such that $\mu_n= <\{A \mathcal{H}^n,B^{\dag}\}> $.
From its moment expansion it is possible to reconstruct the Green's function
as a continued fraction (CF) \cite{Cyrot-Lackmann73}, but
a better conditioned approach is to directly obtain the
CF coefficients from the recursion method of Haydock
\cite{Haydock84}, which can be directly applied in the space
of operators, with the Liouvillian playing the role of the
Hamiltonian in the standard RP. The more conventional wave-function RP
consists in starting from an initial chosen state
$\psi_0$ and then generating a set of orthonormal states $\psi_n$ ($n>0$)
by the recurrence relation ($b_0\equiv0$):
  \beq
  H \psi_n= a_n \psi_n+b_n \psi_{n-1}+b_{n+1} \psi_{n+1}
  \eeq
 The coefficients $a_n$ and $b_{n+1}$  are obtained by
 the scalar product of $\psi_n$ by $H \psi_n$  and  with the norm
  $\parallel H \psi_n - a_n \psi_n-b_n \psi_{n-1}\parallel$ respectively.
In the $\psi_n$ basis, the Hamiltonian has a tridiagonal form,
which can be represented as a semi-infinite tight-binding (TB) chain.
Once a sufficient number of coefficients pairs have been calculated, the
diagonal element of the resolvent $(\omega - H)^{-1}$ can be expressed
as a CF. Standard techniques exist to terminate this expansion \cite{TurchiDucastelle}.
The application of this approach to operator space is
straightforward, each vector now corresponding with an operator.
For our purposes we will be mainly concerned with the
local Green function, where $A=B=c_0$, the destruction operator of
an electron of given spin at site $0$. The action of the
Liouvillian on any creation or destruction operator (or any
combination of them) is then easily computable from its definition,
Eq.~(\ref{liouville}). A natural choice for the scalar product, which is
necessary to compute the CF coefficients, is $(A|B)=
 <\{A,B^{\dag}\}>$ as suggested by Eq.~(\ref{momentexp}), where
thermal averages reduce to their ground-state expectation value at zero
temperature. Details for computing them will be given in
a forthcoming paper.
In this formalism, the Green's function appears as the diagonal element of
the resolvent of the Liouvillian:
  \beq
   G(\omega)\equiv <<c_0;c_0>>_{\omega}=(c_0(\omega-\mathcal{H})^{-1}|c_0)
\eeq
This expression is the origin of the name we have proposed for this approach:
the Extended Recursion in Operator Space
(EROS); it reduces to the usual RP when only one-body operators appear in the Hamiltonian.

We now apply it to the SIAM Hamiltonian:
\beqn H_{SIAM}&=&\epsilon_0 \sum_{\sigma}n_{0 \sigma}+U
n_{0 \uparrow}n_{0 \downarrow}  \nonumber\\
  + \sum_{k \sigma} \varepsilon_k c_{k
\sigma}^\dagger c_{k \sigma}
 &+& \sum_{k  \sigma}(V_{k \sigma}c_{0 \sigma
}^\dagger c_{k \sigma} + V_{k \sigma}^{*}c_{k \sigma}^\dagger c_{0
\sigma}),
  \label{SIAM}\eeqn  which describes a localized  orbital $0$ that has
electronic correlations because of the Coulomb
interaction $U$ and is coupled to itinerant non-interacting
electrons of dispersion $\varepsilon_k$; the latter is reflected by an
hybridization function $\Delta(\omega)= \sum_k
\frac{|V_k|^2}{\omega-\varepsilon_k}$, when spin indices are ignored.
\begin{figure}[h!]
\centerline{\psfig{figure=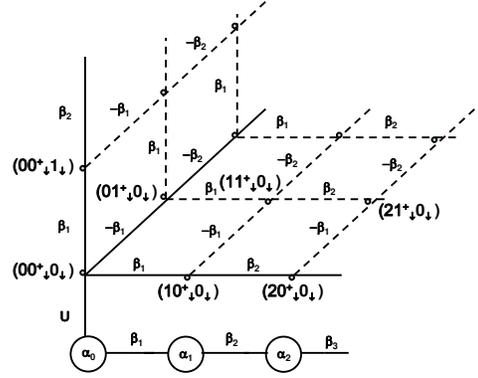,width=8cm,angle=0,scale=1}}
\vspace{-1cm}
 \caption{\label{fig0}Schematic representation of
the Liouvillian}
\end{figure}
If we use a conventional RP to tridiagonalize  the last 2 terms, we
can rewrite Eq.~(\ref{SIAM}) as:
\beqn H^{imp}= U
n_{0\uparrow}n_{0\downarrow}+ \sum_{\sigma } \epsilon_0
n_{0\sigma}\nonumber\\+\sum_{p>0 \sigma } ( \alpha_p c_{p
\sigma}^\dagger c_{p \sigma}+ \beta_p c_{p-1 \sigma}^\dagger c_{p
\sigma}+ \beta_{p+1} c_{p+1 \sigma}^\dagger c_{p \sigma}
)\label{Himp1}\eeqn
These coefficients $\{\alpha_p,\beta_{p+1}\}$, are the
tight-binding parameters of a semi-infinite chain and also
those of the CF expansion considered as a parametrization of the
hybridization function $\Delta(\omega)$, as already often noticed
in NRG context
\cite{Uhrig04}), whereas a RP applied to Hamiltonian (\ref{Himp1})
provides the CF coefficients $\{a _n,b_{n+1}\}$ of the impurity
Green function $G(\omega)$. Eqs.~(\ref{liouville}) and (\ref{Himp1})
give the action of $\mathcal{H}$
on the basis operators that is necessary to perform an operator RP (where the up-spin
is assumed if a spin-index is omitted):
\beqn c_0 \mathcal{H}=\varepsilon_0 c_0+\beta_1 c_1+
U c^{}_0 c^{\dagger}_{0\downarrow} c^{}_{0\downarrow}\\
c_p \mathcal{H}=\alpha_p c_p+\beta_p c_{p-1}+\beta_{p+1} c_{p+1}
\;\; p>0. \eeqn
Compared to the ``natural" operators $c_p$, related
to a given state $p$, the local 2-body interaction
$\mathcal{H}_{int}$ due to $U n_{0\uparrow}n_{0\downarrow}$ has
generated a new operator $c^{}_0 c^{\dagger}_{0\downarrow}
c^{}_{0\downarrow}$, which is the origin of a site representation
for 1/8 of a simple
tri-dimensional cubic lattice, where each site represents an
operator $c^{}_p c^{\dagger}_{q \downarrow} c^{}_{r \downarrow}$ indexed
by 3 positive integers $p, q, r$ (see Fig. \ref{fig0}). This
construction is supported by the following observation, which can be
systematically extended:
\beqn c_0 n_{0\downarrow}\mathcal{H}=Uc_0
n_{0\downarrow}+
\beta_1 c^{}_1 c^{\dagger}_{0 \downarrow} c^{}_{0 \downarrow}
-\beta_1 c^{}_0
c^{\dagger}_{1 \downarrow} c^{}_{0 \downarrow}\nonumber\\
+\beta_1 c^{}_0 c^{\dagger}_{0 \downarrow} c^{}_{1 \downarrow} \eeqn
The RP can be performed in this landscape as easily as the regular
recursion on an usual lattice through knowing the action of $\mathcal{H}$
on operators like
$c^{}_p c^{\dagger}_{q \downarrow} c^{}_{r \downarrow}$.
\begin{figure}[h!]
\vspace{-2cm}
\centerline{\psfig{figure=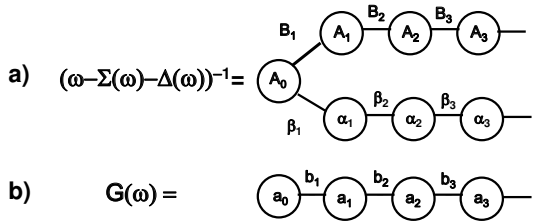,width=8cm,angle=0,scale=1.2}}
\vspace{-2cm} \caption{\label{fig2} Two different recursion processes for $G(\omega)$ }
\end{figure}
Special attention should be paid for the case where one or
two of the indexes $p,q,r$ are zero, viz., the 3 faces and the 3
edges of the 1/8 of the cubic lattice, since in these cases
$\mathcal{H}_{int}$ has a non-zero effect. For the edges, this operation gives
either $0$ or $U$ for the on-site term, but for the faces, a new kind of
operator is generated, which is the product of 5 ``natural"
operators (3 destruction and 2 creation operators), and
the action of the Liouvillian (which are not representable by Fig.~\ref{fig0})
can be represented as the
sites of 5-dimensional hypercube.
These operators correspond to the motion of a hole
with 2 electron-hole pairs.  In our calculations we did not take them exactly into
account, but instead projected them onto existing products of
3 operators (which corresponds to a certain EOM decoupling) or
simply neglected them; in term of moments their influence only starts with the 7th moment.

The self-energy, $\Sigma(\omega)$, which is related to the Green's function through
\beq G(\omega)=(\omega-\Sigma(\omega)-\Delta(\omega))^{-1},
\label{GSigDelta}\eeq
can be expanded as a CF:
 \beqn \Sigma(\omega)&=&A_0+{B_1^2\over\displaystyle \omega-A_1--\cdots}.
\label{SigCF}
\end{eqnarray}
To determine the coefficients of this CF, we note that
the right hand side of Eq.~(\ref{GSigDelta}), which includes the sum of
$\Delta$ and $\Sigma$ in the denominator, can be expanded in a
CF by a RP on their 2 representative chains coupled by their
common first site as shown in Fig.~\ref{fig2}(a) \cite{JTM01};
at the same time this also has to give the
the CF expansion for $G$ in Fig.~\ref{fig2}(b), which is calculated
from the RP in operator space. This enables
the coefficients of Eq.~(\ref{SigCF})
for the self-energy $\Sigma$ to be determined.

We now describe an application of our EROS methodology to a DMFT
solution of the Hubbard model.
An excellent description of the DMFT method is given in \cite{Georges96}
and a recent review \cite{Held07} describes recent developments and extensions.
The Hubbard model on a lattice (for site indices $i$ and $j$) can be written as
\beq H= \sum_{i\neq
j,\sigma} t_{ij} c_{i\sigma}^\dagger c_{j\sigma}+U\sum_i
n_{i\uparrow}n_{i\downarrow} \label{HHub}\eeq The motivation for
DMFT arises from the observation based on a diagrammatic analysis that in infinite dimensions
\cite{MetznerVollhardt89} the self-energy becomes local,
\textit{i.e.}, $k$-independent or simply $\Sigma(\omega)$.
In lower dimensions, this is often only approximately true. In the local limit for $\Sigma$ the
Hubbard model reduces to the problem of a correlated impurity at  site
``$0$" embedded in an effective medium where all the effects of
correlation are represented by the self-energy $\Sigma(\omega)$.
This can be considered as a complex and energy-dependent
on-site energy in terms of a strong analogy with the TB language
used in the Coherent Potential Approximation (CPA,for a detailed discussion see \cite{CPAGonis}), which was developed for studying
alloys and has now been further  extended for strongly correlated electron systems \cite{Kake02}:
\beq H_{DMFT}= U
n_{0\uparrow}n_{0\downarrow}+\sum_{i\neq j,\sigma} t_{ij}
c_{i\sigma}^\dagger c_{j\sigma}+\sum_{i\neq 0 \sigma }
\Sigma(\omega)n_{i\sigma} \label{Himp2}\eeq
It can be shown that this problem
maps onto a SIAM model, Eq.~(\ref{SIAM}). To understand this
within our approach, we start by noticing that the second term
of Eq.~(\ref{Himp2}), the ``hopping" term, can be tridiagonalized
by a RP to provide the CF expansion of the ``bare" hybridization function
$\Delta(\omega)$.  The last term then functions like a constant onsite
energy in a TB Hamiltonian that simply shifts the frequency by $\Sigma(\omega)$.
Hence the last two terms of Eq.~(\ref{Himp2}) can be represented
by a Green's function for an effective TB model with the bare
hybridization function $\Delta(\omega)$ replaced with an effective
$\bar{\Delta}(\omega) = \Delta(\omega-\Sigma(\omega))$.
To complete the DMFT method, we then require that this
self-consistent condition be supplemented by the requirement that the
local impurity Green's function, computed from the Hamiltonian of
Eq.~(\ref{Himp2}), be equal to the effective lattice Green's
function, which has $\Sigma(\omega)$ on all sites, including
site ``$0$". In RP language, this causes the the local $\Sigma(\omega)$
to be added to the on-site terms $\alpha_{p}$ in the
chain representation for $\Delta(\omega)$. Including this
self-energy at site ``$p$" is equivalent to attaching to the TB
semi-infinite chain the CF expansion of $\Sigma$ as parameters
(see Appendix of \cite{JM93}), and leads to a comb-shape topology for
computing $\bar{\Delta}(\omega)$. Having performed
again a RP on this object provides a CF expansion for
$\bar{\Delta}(\omega)$. The DMFT self-consistency is
achieved by simultaneously satisfying all of these CF equations as
one steps through the recursion procedure. The first few
recursion steps generate the CF coefficients $A_0\equiv
U<n_{0-\sigma}>$ of $\Sigma$, corresponding to the Hartree-Fock (HF)
energy-independent self-energy, and then $B_1=Un(n-1)$ and
$A_1=U(1-n)$ gives Hubbard I, which is exact in the atomic limit.
If only operators $c_p c^{\dagger}_{0 \downarrow}
c_{0 \downarrow}$ on one edge of the cubic lattice are retained,
Hubbard III is recovered. Further recursions generate approximations
that go well beyond these (see below). It is worth mentioning that our approach has
some similiarities with the approach developed in
\cite{KakeFulde04}, but our direct use of a RP enables us to go
beyond this work by avoiding their requirement to orthogonalize
each new operator, which becomes becomes difficult after the first few
steps.

\begin{figure}[h!]
\centerline{\psfig{figure=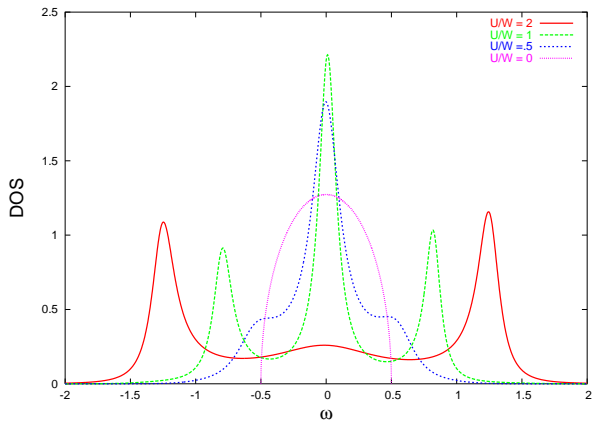,width=8cm,angle=270,scale=0.7}}
\caption{\label{spectDEns.5} RP DMFT calculation of the Hubbard model for the spectral density vs energy for different values of U (color on-line)}
\vspace{-.5cm}
\end{figure}
We have benchmarked our method on the well-studied half-filled case.
For simplicity, all calculations were performed at zero temperature
and used as a HF solution of Eq.~(\ref{Himp2}) as an approximate
ground state for computing the scalar products of the operators.
In the future we will try to use a
better ground state such as a Gutzwiller approximation or an exact
diagonalization for a given number of sites. By including approximate
excited states in a thermal average, finite-temperature calculations are
also possible. For the lattice
model that describes the itinerant electrons, we used
a semi-elliptic non-interacting density of states, with the energy
scale set by the bandwidth. Other lattices would
only change the input CF coefficients for the hybridization
function. For several different values of $U$, the spectral
density is displayed in Fig.~\ref{spectDEns.5}. One clearly
observes the characteristic three-peak features in the Fermi liquid
regime: the coherent contribution of the central peak, and the lower
and upper Hubbard subbands with a splitting of the order of $U$.
One also observes a redistribution of the spectral weight of the coherent central peak
as the interaction $U$ increases.
The Fermi liquid behavior and the
intermediate regime can be better monitored by considering
the real and imaginary parts of the self-energy.
In the Fermi liquid regime, we have checked that the imaginary part of the
self-energy vanishes as it should be at the Fermi level $E_F$, and that
it has the usual $(\omega-E_F)^2$ behavior around $E_F$.

We have presented a new impurity self-consistent solver for the
SIAM model and a DMFT solution of the Hubbard model
using a RP procedure in an operator space. The lowest order
approximations to this method generate the Hartree-Fock, Hubbard I, and Hubbard III
approximations.  Increasing the continued fractions go well beyond these solutions,
generating the coherent Kondo peak, for example, and can provide increasingly
accurate results.
The way that self-consistency is
achieved, step by step, makes the method efficient and
very promising for its use in DMFT approaches and other more
complex correlation problems.

JPJ is grateful to J.X. Zhu (Los Alamos) for helpful discussions, and F. Harris
(Gainesville) for sharing with us his code to compute scalar
product of operators. This work was carried out under the auspices
of the National Nuclear Security Administration of the U.S.
Department of Energy at Los Alamos National Laboratory under
Contract No. DE-AC52-06NA25396. Financial support from
CNLS, Los Alamos and DGA (Delegation Generale pour l'Armement Contract No.
07.60.028.00.470.75.01) are gratefully acknowledged.

\end{document}